\documentclass[useAMS,usenatbib]{mn2e}
\usepackage{graphicx}                   %for PS/EPS graphics inclusion, new
\usepackage{float}
\usepackage{amsmath}
\usepackage{amssymb}
\usepackage{multirow}
\usepackage{upgreek}
\usepackage{bm}
\usepackage{color}
\input{epsf.sty}                        %for PS/EPS graphics inclusion, old
\input{psfig.sty}
\date{}

\begin{document}

\title[Torque reversal during delayed spin up glitch in SGR 1935]{On torque reversal during delayed spin up glitch of SGR 1935+2154}

\author[Wei-Hua Wang et al.]{Wei-Hua Wang $^{1}$\thanks{E-mail: wang-wh@pku.edu.cn},
Ming-Yu Ge$^{2}$, Xi Huang$^{3}$ and Xiao-Ping Zheng$^{4,5}$\thanks{E-mail: zhxp@ccnu.edu.cn}\\
% List of institutions
$^1$College of Mathematics and Physics, Wenzhou University, Wenzhou 325035, China\\
$^2$Key Laboratory of Particle Astrophysics, Institute of High Energy Physics, Chinese Academy of Sciences, Beijing 100049, China\\
$^3$School of Mathematical and Physical Sciences, Wuhan Textile University, Wuhan 430200, China\\
$^4$Institute of Astrophysics, Central China Normal University, Wuhan 430079, China\\
$^5$School of Physics, Huazhong University of Science and Technology, Wuhan 430079, China\\
}

\maketitle

\begin{abstract}
Ge et al. reported a peculiar large glitch observed from SGR 1935 recently~\citep{2022arXiv221103246G}.
Interestingly, this glitch occurred about $3.1\pm 2.5~\rm{days}$ before FRB 200428, accompanied by a delayed spin up (DSU) process with a timescale of $8\pm 1~\rm{days}$.
This DSU is the first one detected from magnetars unambiguously, previously, DSU has only been detected from large glitches of the young Crab pulsar.
Strikingly, this DSU even resulted in torque reversal, i.e., the spin down state of SGR 1935 was turned into a spin up one within $8\pm 1~\rm{days}$, the first case observed from isolated neutron stars as far as we know.
What's more, the average positive torque is consistent in value and order of magnitude with those deduced from DSUs observed from the Crab pulsar.
The coincidences of DSU appearance and consistent positive torque in the Crab pulsar and SGR 1935 suggest that, these DSUs have the same physical origin and should be analyzed simultaneously.
Joint analysis on DSUs from the Crab pulsar and SGR 1935 suggests a physical origin independent of (at least not directly dependent on) rotational parameters.
If this new DSU and torque reversal phenomena be further confirmed and accretion be excluded convincingly, nearly all models regarding DSU phenomenon under the framework of vortex motion should be reconsidered, new physics will be required.
We stress that, joint analysis on DSUs from different pulsars with different rotational parameters may provide pivotal clue to reveal their physical origin, even neutron star equation of state and the triggering mechanism(s) of (galactic) fast radio bursts ultimately.
\end{abstract}

\begin{keywords}
stars: neutron-stars: magnetars-pulsars: general-pulsars: individual: PSR B0531+21, SGR 1935+2154
\end{keywords}

\section{Introduction}
Recently, Ge et al. reported that, the timing behaviors around the fast radio burst (FRB) 200428 from the soft gamma ray repeater (SGR) 1935+2154 (SGR 1935 for short hereafter) are very typical for a glitch, including the persistent and delayed spin up and the decaying components~\citep{2022arXiv221103246G}.
According to their work, relative size of this glitch is $\Delta\nu/\nu\sim 6\times 10^{-5}$.
Interestingly, this glitch occurred about $3.1\pm2.5~\rm{days}$ before FRB 200428,
possibly indicating some physical associations between glitch and FRB-like bursts.
What's more, this glitch contains a 8-day-long slow rise component, similar with the delayed spin up process (DSU) observed in large Crab pulsar glitches.
Specifically, the occurrence of DSU turns its spin down state into a spin up one (torque reversal hereafter), which means the existence of an additional positive torque dominating over the original braking torque.
The positive torque decays exponentially within a timescale of
about $8~\rm{days}$, after which SGR 1935 spins down again.
We note that, SGR 1935 is the second pulsar where DSU has been observed.
The glitch associated torque reversal is even more interesting.
Torque reversal and delayed spin up are the focus of this work.

Glitches are characterized by abrupt increases in the spin frequency of pulsars, followed by the relaxation process towards the pre-glitch state.
Most of the glitches are accompanied by sudden increases in the spin down rates ($\Delta\dot\nu<0$).
As of this writing, more than $600$ glitches have been detected in $208$ pulsars~\footnote{Detailed information on publicly reported glitches can be found at http://www.jb.man.ac.uk/pulsar/glitches/gTable.html~\citep{2011MNRAS.414.1679E}}.
For rotation-powered pulsars, $\Delta\dot\nu/\dot\nu$ at the glitch epoch lies in the range $10^{-4}-10^{-2}$, i.e., the braking enhanced slightly following glitches (the spin down rate $\dot\nu<0$).
While for high magnetic field pulsars and magnetars, $\Delta\dot\nu/\dot\nu>1$ cases have also been found~\citep{2018MNRAS.480.3584D} or deduced~\citep{2020MNRAS.497.2680T}.
In some cases, $\Delta\dot\nu/\dot\nu<0$, which means the braking decreased following glitches.
If $\Delta\dot\nu/\dot\nu<-1$, which means $\Delta\dot\nu+\dot\nu>0$, the torque will be reversed.
However, before this glitch in SGR 1935, glitch associated torque reversal has never been observed from isolated neutron stars (NSs) as far as we know.

\begin{table*}
\begin{center}
    \caption{Observational summary of seven delayed spin-up events}
    \begin{tabular}{ | p{1.5cm} | p{0.4cm} | p{2.4cm} |p{1.6cm}| p{1.3cm}|  p{1.3cm}| p{2.0cm}| p{1.2cm}|}
    \hline\hline
    Pulsar  &Year &Glitch epoch  &$(\Delta\nu/\nu)~\rm{[\mu Hz]}$  &$\Delta\nu_{d}~\rm{[\mu Hz]}$    &$\tau_{d}~\rm{[days]}$     &$\Delta\dot\nu_{d}/(10^{-12}~{\rm{s^{-2}}})$   & Refs.$^a$\\
    \hline
    J0534+2200   &1989  &MJD 47767.504(3)   &$\sim2.35$         &$0.7$            &$\sim 0.8$     &$\sim10.1$    &(1)\\
                         &1996  &MJD 50260.031(4)   &$\sim0.95$         &$0.31$           &$\sim 0.5$     &$\sim7.18$    &(2)\\
                         &2004  &MJD 53067.0780(2)  &$\sim6.37$         &$0.35(5)$        &$1.7(8)$*      &$\sim2.38$*   &(3)\\
                         &2011  &MJD 55875.49(3)    &$\sim1.16$         &$0.43(5)$        &$1.6(4)$*      &$\sim3.1$*    &(3)\\
                         &2017  &MJD 58064.555(3)   &$15.3$             &$1.071(4)$       &$1.703(13)$    &$\sim7.27$    &(4)\\
                         &2019  &MJD 58687/565(4)   &$0.94(3)$          &$0.343(18)$      &$0.76(7)$      &$\sim5.22$    &(5)\\
    \hline
    SGR 1935             &2020 &MJD 58964.5(2.5)   &$19.8(1.4)$        &$5.88(9)$        &$8(1)$         &$\sim8.51$    &(6)\\
    \hline\hline
    \end{tabular}
    \label{delayed spin up parameters}
\end{center}
Notes:
$^*$Results taken from X-ray observations.\\
$^a$ \scriptsize{References: (1)~\cite{1992Natur.359..706L};(2)~\cite{2001ApJ...548..447W};(3)~\cite{2020ApJ...896...55G};(4)~\cite{2018MNRAS.478.3832S};(5)~\cite{2021MNRAS.505L...6S};(6)~\citep{2022arXiv221103246G}} \\
The number in bracket represents errorbar of the last significant digit.\\
\end{table*}

Theoretically, glitch is widely accepted as a probe of NS physics, such as NS equation of state~\citep{1999PhRvL..83.3362L,2015MNRAS.454.4400N,2020MNRAS.492.4837M} or NS mass~\citep{2015SciA....1E0578H,2017NatAs...1E.134P}.
Glitches are believed to be a manifestation of sudden vortex outward motion and angular momentum transfer from crustal superfluid to the crust (and that coupled to it)~\citep{1975Natur.256...25A, 2011ApJ...743L..20P}, whether the outer core superfluid get involved or not is still under debate.
Based on aforementioned framework, Alpar et al. established the vortex creep model to describe the post-glitch recovery process and achieved great success~\citep{1984ApJ...276..325A}.
Some improved models further considered the hydrodynamical process of vortex motion, trying to extract information (such as the mutual friction) through glitch rise process~\citep{2018MNRAS.481L.146H, 2018ApJ...865...23G, 2020MNRAS.493L..98S}.

Observationaly, no glitch rise has been resolved completely till now.
Previously, rise timescale of 2000 Vela glitch was estimated to be $~40~\rm{s}$~\citep{2002ApJ...564L..85D}.
Recently, rise timescale of 2017 Vela glitch was constrained to be $~48~\rm{s}$~\citep{2018Natur.556..219P}, Ashton et al. further constrained it to be less than $12.6~\rm{s}$ with $90\%$ confidence level~\citep{2019NatAs...3.1143A}.
Apart from these, several Crab glitches have been partially resolved in time, and a slow rise process with timescale of days have been resolved~\citep{2021MNRAS.505L...6S}.
The slow rise phenomenon was named delayed increases in rotation rate~\citep{1992Natur.359..706L}, delayed spin up~\citep{2018MNRAS.478.3832S} or extended spin up~\citep{2021MNRAS.505L...6S} in different papers previously.
We call the slow rise phenomenon as {\it{delayed spin up (DSU)}} hereafter, glitches which contain DSU are called {\it{delayed spin up glitch}} in this work.

DSU is a day-long-timescale slow rise phenomenon firstly discovered during the 1989 Crab pulsar glitch~\citep{1992Natur.359..706L}.
Interestingly, in the following three decades, DSU has only been observed from the Crab pulsar.
Apart from this, six out of the top seven Crab glitches (in size) are found to be accompanied by DSU, suggesting that DSU does not occur occasionally.
The other five DSUs were found in the large Crab pulsar glitches of 1996~\citep{2001ApJ...548..447W}, 2004, 2011~\citep{2020ApJ...896...55G}, 2017~\citep{2018MNRAS.478.3832S} and 2019~\citep{2021MNRAS.505L...6S}.
The exception is the large Crab glitch of 1975 (MJD 42447.26), around which no high cadence of observations were available.
Details of all the currently known DSUs, including the latest one discovered by Ge et al., are present in Table {\ref{delayed spin up parameters}}.

Discovery of DSU from SGR 1935 has enlarged the DSU population, what's more, a joint analysis on DSU observed from SGR 1935 and the Crab pulsar is now possible, which allows to test current DSU models and helps to shed light upon the physical origins behind DSUs and the torque reversal phenomena.
In this work, We focus on all the DSU events and the torque reversal phenomenon observed from SGR 1935, trying to reveal their theoretical indications.
The glitch/spin-down glitch and FRB associations will also be discussed briefly.

This article is organized as follows.
We present rotational parameters of SGR1935 and details of its peculiar glitch before FRB 200428 in section {\ref{sec2}, we present a joint analysis on DSU events from the Crab pulsar and SGR 1935 in section {\ref{sec3}.
In section {\ref{sec4}, we discuss the theoretical indications of torque reversal during DSU glitch before FRB 200428 observed from SGR 1935.
Finally, the conclusions and discussions are present in section {\ref{sec5}.

\section{Rotational parameters of SGR 1935 and its peculiar glitch}
\label{sec2}
SGR 1935 is an isolated neutron star with a spinning frequency $\nu=0.308~\rm{Hz}$ and a spin down rate $\dot\nu=-1.356\times 10^{-12}~\rm{s^{-2}}$, implying a surface dipole magnetic field $B\sim 2.2\times 10^{14}~\rm{G}$ and a characteristic age
$\tau_{\rm c}=3.6~\rm{kyr}$~\citep{2016MNRAS.457.3448I}.
According to its features such as spin period, inferred ultra strong magnetic field strength and prolific X-ray flaring behaviors,  SGR 1935 was confirmed to be a magnetar~\citep{2016MNRAS.457.3448I}.
SGR 1935 emitted hundreds of short bursts between April and May 2020,
among which one non-thermal double-peaked X-ray burst is coincident with FRB 200428~\citep{2021NatAs...5..378L}.
Ever since the discovery of FRB 200428, SGR 1935 has been intensely monitored.

Utilizing observations from {\it{NICER}}, {\it{NuSTAR}} and XMM-Newton, Ge et al. found a glitch from SGR 1935 before FRB 200428~\citep{2022arXiv221103246G}.
This glitch occurred $3.1\pm2.5~\rm{days}$ before FRB 200428, with glitch size $\Delta\nu/\nu\sim 6\times 10^{-5}$ ($\Delta\nu=19.8\pm1.4\rm{\mu Hz}$) and $\Delta\dot\nu/\dot\nu=-4.4\pm0.7$ (torque reversal) at the glitch epoch.
The glitch rise can be divided into two parts, the unresolved glitch rise and the resolved delayed spin up.
Frequency increase during DSU is $\Delta\nu_{d}=5.88\pm 0.09~\rm{\mu Hz}$, timescale of this DSU is $\tau_{d}=8\pm 1~\rm{days}$.
The DSU and torque reversal are what we are focusing on in this work.
For more details about this glitch, please turn to panels (d) and (e) in Figure 1 and Table 2 in~\cite{2022arXiv221103246G}.

\section{Torque reversal and delayed spin up}
\label{sec3}
We perform a joint analysis on the DSU events from the Crab pulsar and SGR 1935 here.
In the standard vortex model, glitch occurs when crustal vortices unpin from the pinning sites abruptly, this process is accompanied by unresolved fast glitch rise (due to sudden unpinning and angular momentum transfer) and enhanced braking (due to superfluid decoupling) at the glitch epoch, i.e., $\Delta\dot\nu/\dot\nu>0$, the braking enhanced by $10^{-4}-10^{-2}$ in most cases~\footnote{http://www.jb.man.ac.uk/pulsar/glitches/gTable.html~\citep{2011MNRAS.414.1679E}}.

However, DSU represents a slow glitch rise, which corresponds to the existence of an additional positive torque.
Average value of the positive torque is $\Delta\dot\nu_{d}=\Delta\nu_{d}/\tau_{d}$.
The $1989$, $1996$, $2017$ and $2019$ DSUs observed from the Crab pulsar through radio observations give $\Delta\dot\nu_{d}\sim (5.22-10.1)\times 10^{-12}~{\rm{s^{-2}}}$, which amounts to $1.42-2.75$ percent of its steady spin down rate ($\nu_{\rm crab}=29.58~\rm{Hz}$, $\dot\nu_{\rm{crab}}=-3.67\times 10^{-10}~{\rm{s^{-2}}}$)~\footnote{http://www.jb.man.ac.uk/~pulsar/crab.html \citep{1993MNRAS.265.1003L}}.
The $2004$ and $2011$ DSUs observed from the Crab pulsar through X-ray observations resulted in $\Delta\dot\nu_{d}\sim (2.38-3.1)\times 10^{-12}~{\rm{s^{-2}}}$, slightly smaller than that deduced from radio observations, but remains consistent in order of magnitude.
Most interestingly, the $2020$ DSU observed from SGR 1935 through X-ray observations gives $\Delta\dot\nu_{d}\sim 8.51\times 10^{-12}~{\rm{s^{-2}}}$, consistent in value and order of magnitude with that in the Crab pulsar.
What's more, this positive torque even exceeded the braking torque of SGR 1935, i.e.,a net positive torque reached. It means that the appearance of DSU in SGR 1935 turned its spin down state into a spin up one for a period within $\tau_{d}$.

The consistency in value and order of magnitude of the additional positive torque between the Crab pulsar and SGR 1935 is suggestive.
Spin frequency of the Crab pulsar is $\nu=29.58~\rm{Hz}$, $96$ times that of SGR 1935.
Spin down rate of the Crab pulsar is $\dot\nu_{\rm{Crab}}=-3.67\times 10^{-10}~{\rm{s^{-2}}}$, $270$ times that of SGR 1935.
The vast differences in rotational parameters and the consistency in additional positive torque between the Crab pulsar and SGR 1935, and the appearance of torque reversal in SGR 1935 suggesting {\it{no direct association between DSU and rotational parameters}}.

\section{Theoretical indications}
\label{sec4}
We discussion the physical origins of delayed spin up and associated torque reversal in literature in this section.

\subsection{External origins}
Short-timescale accretion process may provide the additional positive torque, resulting in the torque reversal (e.g.,~\cite{2022MNRAS.517L.111L}) and DSU phenomena.

The possibility of accretion in SGR 1935 is low from two aspects.
Firstly, the isolation nature of SGR 1935 does not support the accretion scenario.
Secondly, if accretion exists in SGR 1935, the sudden appearance of additional positive torque and its exponential decay indicate the accretion operates abruptly and decay quickly.
This seems to be unreasonable as accretion usually proceeds continuously.
We stress that, to confirm the non-existence of accretion convincingly, dedicated analysis on the pre-glitch and post-glitch spectrums needs to be performed.

For the Crab pulsar, radio and X-ray observations show no flux enhancement or pulse profile change around the largest Crab glitch~\citep{2018MNRAS.478.3832S,2020A&A...633A..57V,2022ApJ...932...11Z}.
These observations suggest the internal origin of DSU in the Crab pulsar.

\subsection{Internal origins}
There have been several models proposed to explain the DSU phenomenon from internal interplay view.
According to the difference in physical mechanisms, these models are divided into two groups, the superfluid vortex creep models~\citep{1984ApJ...276..325A,1996ApJ...459..706A,2015MNRAS.449..933A,2018MNRAS.481L.146H,2019MNRAS.488.2275G} and the excess heating models~\citep{1979ApJ...231..880G,1996ApJ...457..844L}.

\subsubsection{DSU in the superfluid vortex creep models}
Alpar et al. attributed the positive torque during DSU to the response of the vortex creep process to a component of the glitch involving vortex inward motion~\citep{1996ApJ...459..706A,2015MNRAS.449..933A,2019MNRAS.488.2275G}.
In the standard vortex model, the initial increase in spin down rate (additional negative torque) corresponds to superfluid decoupling~\citep{1984ApJ...276..325A}, i.e., the external torque acts on less moment of inertia instantly after the glitch.
After the angular momentum transfer which takes less than a minute~\citep{2018ApJ...865...23G}, response of the outward vortex creep process to different glitch involving components results in the additional negative torque observed in post-glitch process~\citep{1984ApJ...276..325A,1989ApJ...346..823A}.

In order to explain the DSU in 1989 Crab glitch, Alpar et al. proposed the notion of newly formed vortex traps occurred during a quake, and attributed the anomalous DSU to the response of vortex creep to this region~\citep{1996ApJ...459..706A}.
Later, the notion of newly formed vortex traps was replaced by a crustal plate which breaks and moves towards the rotation axis~\citep{2015MNRAS.449..933A,2019MNRAS.488.2275G}.
Anyway, both the newly formed vortex trap region and crust breaking and moving towards the rotation axis scenario will result in vortex inward motion, which correspond to a positive torque during post-glitch recovery process.
According to equation (12) in~\cite{2019MNRAS.488.2275G}, the positive torque $\Delta\dot\nu_{d}$ has the form
\begin{align}
\Delta\dot\nu_{d} (t)=
&-\frac{I_{0}}{I}\left|\dot\nu\right|\left[1-\frac{1}{1+\left({\rm{e}}^{-t'_{0}/\tau'_{\rm nl}}-1\right){\rm{e}}^{-t/\tau'_{\rm nl}}}\right]\propto\frac{I_{0}}{I}\left|\dot\nu\right|,
\end{align}
all the above parameters follow their definitions.
Apparently, this positive torque is dependent on the spin down rate $\dot\nu$.
Most importantly, this torque will never exceed $\left|\dot\nu\right|$, as it is impossible for the whole star to get involved in inward vortex motion, i.e., $I_{0}<I$.
Actually, according to their fitting, $I_{0}/I\sim 1.5\times 10^{-3}$ for the 2017 Crab glitch~\citep{2019MNRAS.488.2275G}, much less than unity.
Therefore, the positive torque associated with inward vortex motion is insufficient to explain the DSU and torque reversal observed from SGR 1935.

Haskell et al. present a two-component model for DSU recently.
They take both the rise of the Vela pulsar glitches and DSU of the Crab pulsar glitches as the consequences of hydrodynamical process of vortex motion (equation (1) in their work), and attribute the different rise timescales in the Vela and Crab pulsars to the difference in mutual friction in locations where glitches are triggered~\citep{2018MNRAS.481L.146H}.
In order to reproduce the day-long timescale DSU in the Crab pulsar, Haskell et al. assumed the existence of vortex sheet (a region with high vortex density) and large spin lag, which helps to prolong the timescale of vortex outward motion.
However, as illustrated in right panels in Figure 1 in their work, their non-linear simulations results in $\Delta\nu$ increases (almost) linearly in the DSU stage, which corresponds to a constant additional positive torque, inconsistent with the days-long timescale positive torque evolution as shown in panel (D) in Figure 1~\citep{2018MNRAS.478.3832S} and panel (e) in Figure 1~\citep{2022arXiv221103246G}.
Even though the torque evolution in the Crab pulsar can be reproduced by adjusting parameters, both the evolution equation for vortex motion and free vortex fraction have dependency on rotational parameters (see equations (1) and (8) in their work).
For DSU in the Crab pulsar, the average positive amounts to several percent of its spin down rate.
Therefore, due to the rotational dependency, it is reasonable to suspect that, the corresponding positive torque in this model will not exceed the braking torque of SGR 1935, i.e.,there is no possibility of torque reversal.

Sourie \& Chamel recently developed a three-component model and showed that, the pinning of vortices to fluxoids could affect the mutual friction parameter and give rise to delayed spin up events~\citep{2020MNRAS.493L..98S}.
Previously, Graber et al. developed a three-component model to monitor the glitch rise process, but this model can not reproduce the DSU as the glitch rise timescale is no more than minutes even for extremely small mutual friction parameter~\citep{2018ApJ...865...23G}.
In this new work, both the overshoot in the vela pulsar and DSU in the Crab pulsar can be reproduced by varying the number of fluxoids $N_{{\rm{p}}}$ attached to vortices manually~\citep{2020MNRAS.493L..98S}.
However, the most serious problem of this paper lies in that, the choices of different $N_{{\rm{p}}}$ and thus mutual friction are nonphysical.
What's more, the equations of vortex motion are rotational dependent, the possibility of torque reversal in SGR 1935 is suspected.
If the torque reversal can be reproduced by adjusting mutual friction manually, comparison between the adopted mutual friction and that from dynamic calculations will serve as a test on this model.

\subsubsection{DSU in the excess heating models}
Greenstein proposed the scenario that, an increase in temperature of the star leads to an increase in the frictional coupling between the superfluid and the normal components, response of a superfluid interior to the perturbation in temperature could account for timing behaviors such as slow glitch rise and timing noise~\citep{1979ApJ...231..880G}.
This idea seems to be interesting and plausible, but there exists three problems.
Firstly, glitch in this scenario has no fast rise within timescale of seconds.
Secondly, the required temperature increase seems to be unrealistically large.
Thirdly, if slow rise occurs, the frequency increase during the slow rise stage tends to be much larger than that in the fast rise process, which is actually in contradiction with observations from the Crab pulsar and SGR 1935 (see Table~\ref{delayed spin up parameters}).

Later in 1996, Link \& Epstein proposed the thermal driven picture to account for glitches, similar with that of Greenstein but were different in details~\citep{1996ApJ...457..844L}.
In spite of their merits in reproducing the various glitch behaviours seen in Crab and and Vela pulsar, we have to stress that, this model was constructed under the two-component model, similar with that of Alpar~\citep{1984ApJ...282..533A,1984ApJ...276..325A}.
In this case, almost all the moment of inertia was coupled with the external torque except the weakly pinned crustal superfluid component.
Therefore, it seems difficult to reproduce torque reversal.
Otherwise, torque reversal should be very common in magnetars due to their prolific burst behaviors and strong correlation between large radiative changes and glitch~\citep{2014ApJ...784...37D}.

\section{Conclusions and Discussions}
\label{sec5}

Most previous works focus on the frequency increase ($\Delta\nu_{d}$) during delayed spin-ups, with the presence of more details about DSUs, especially the discovery of torque reversal in SGR 1935, we suggest more attention be paid to how the delayed spin-up proceed ($\dot\nu$), i.e., to the {\it{dynamical process}}.

In this work, we have collected all delayed spin-ups observed from the Crab pulsar and SGR 1935, we presented a simple analysis on the similarities between all these events.
We found a similar additional positive torque between DSUs from the Crab puslar and SGR 1935, this positive torque is even larger than the nominal spin down rate of SGR 1935, i.e., braking torque of SGR 1935 was reversed within $8\pm 1~\rm{days}$.
The great differences in spin frequency and spin down rate of the Crab pulsar and SGR 1935, and the consistency in value and order of magnitude of the additional positive torque indicate the same physical origin rather than coincidence.
If true, the physical process behind should not directly dependent on, or be insensitive to rotational parameters.

Model constraints are present in this work, based on the idea that DSUs from the Crab puslar and SGR 1935 may have the same physical origin.
Special attention was paid to the torque reversal phenomenon during DSU in SGR 1935.
We found that, all models regarding DSUs are essentially based on vortex motion theory in two-component or three-component neutron stars.
Nearly all models are rotational dependent.
Models such as the vortex inward motion in two-component neutron stars will not reproduce torque reversal explicitly~\citep{1984ApJ...276..325A,2019MNRAS.488.2275G}.
Vortex pinning in the core of three-component neutron stars may reproduce features of DSU in the Crab pulsar~\citep{2020MNRAS.493L..98S}, but we are not sure whether torque reversal is possible in SGR 1935 in this model.
Besides, they have to make it clear the physical meaning of mutual friction parameter they chose.
Anyway, all these models are problematic to some extent and should be reconsidered.
Therefore, we suggest joint analysis on DSUs from different pulsars.

Conceptually speaking, we suggest that the reverse torque still comes from the release of superfluid angular momentum, similar with glitch itself.
This is most likely as the angular momentum reservoir is vast in neutron stars.
Besides, superfluid angular momentum transfer is efficient enough to match the sudden appearance of positive torque.
However, the triggering mechanism of superfluid angular momentum release behind DSU shouldn't be the same as that of glitch.

The temporal coincidence between glitch and FRB 200428 possibly suggest some physical connections between them, which is not discussed here and beyond the scope of this work.
However, Younes et al. reported an X-ray silent spin-down glitch about $3$ days before another three moderately bright FRB-like bursts from the same pulsar~\citep{2022arXiv221011518Y}.
The $3$ days time interval is consistent with the $3.1\pm 2.5~\rm{days}$ interval reported in~\cite{2022arXiv221103246G}, is this another coincidence?
Recently, Wang et al. reported the GECAM and HEBS detection of a short X-ray burst (ATel\#15682~\footnote{https://www.astronomerstelegram.org/?read=15682}) associated with another bright radio burst from SGR 1935 detected by CHIME (ATel\#15681~\footnote{https://www.astronomerstelegram.org/?read=15681}).
Based on the two cases above, it is very hopeful that another glitch/spin-down glitch may be associated with this FRB.
We are expecting to check the existence and the size of glitch/spin-down glitch before this new FRB, rise timescale of possible DSU, importantly, the time interval between the glitch and FRB epoch.
If the time interval is consistent with the previous two cases again, these events will strength our idea that, the glitch/spin-down glitch and FRB associations may correspond to an unified picture.
In this case, research on glitch, DSU, torque reversal and spin-down glitch from SGR 1935 may shed light on not only neutron star interior physics, but also the triggering mechanisms of FRBs (or FRB-like bursts)~\citep{2021MNRAS.507.2208W}.

\section*{Acknowledgements}
This work is supported by the National SKA Program of China (Grant No. 2020SKA0120300) and the National Natural Science Foundation of China (Grant No. 12033001), the International Partnership Program of Chinese Academy of Sciences (Grant No.113111KYSB20190020), and Scientific Research Project Fund of Hubei Provincial Department of Education (Grant No. Q20161604).

\section*{Data availability}
The data underlying this work are available in this article, no new data were generated or analyzed.

%\newpage

\end{document}